\documentclass[aps,prl,twocolumn,showpacs,preprintnumbers,amsmath,amssymb]{revtex4-1}
 \def\ep{{\epsilon}}

 \def\frac#1#2{{#1\over #2}}

 \def\s{\sqrt}

\def\be{\begin{equation}}
\def\ee{\end{equation}}
\def\ba{\begin{eqnarray}}
\def\ea{\end{eqnarray}}

 \def\de{\partial}

 \def\lr{\leftrightarrow}
 \def\f {\frac}
 \def\ti{\tilde}
 \def\ap{\alpha}

 \def\no{\nonumber \\}

 \def\la{\langle}
 \def\lb{\rangle}
 \def\ep{\epsilon}

\usepackage{color}
\usepackage{graphicx}% Include figure files
\usepackage{dcolumn}% Align table columns on decimal point
\usepackage{bm}% bold math

\begin{document}

\title{cMERA as Surface/State Correspondence in AdS/CFT}
IPMU15-0077; YITP-15-46
\author{Masamichi Miyaji$^{a}$, Tokiro Numasawa$^{a}$, Noburo Shiba$^{a}$,  \\
Tadashi Takayanagi$^{a,b}$ and Kento Watanabe$^{a}$}

\affiliation{$^a$Yukawa Institute for Theoretical Physics,
Kyoto University, \\
Kitashirakawa Oiwakecho, Sakyo-ku, Kyoto 606-8502, Japan}

\affiliation{$^{b}$Kavli Institute for the Physics and Mathematics
 of the Universe,\\
University of Tokyo, Kashiwa, Chiba 277-8582, Japan}

\date{\today}

\begin{abstract}
We present how the surface/state correspondence, conjectured in arXiv:1503.03542, works in the setup of AdS$_3/$CFT$_2$ by generalizing the formulation of cMERA. The boundary states in conformal field theories play a crucial role in our formulation and the bulk diffeomorphism is naturally taken into account. We give an identification of bulk local operators which reproduces correct scalar field solutions on AdS$_3$ and bulk scalar propagators. We also calculate the information metric for a locally excited state and show that it is given by that of 2d hyperbolic manifold, which is argued to describe the time slice of AdS$_3$.
\end{abstract}

%\pacs{72.10.-d,73.21.-b,73.50.Fq}
\maketitle

{\bf 1. Introduction}

Even though the idea of AdS/CFT correspondence \cite{Ma} has lead tremendous progresses in string theory, we still do not fully know its basic mechanism how it works. It is obvious that the AdS/CFT correspondence can be understood in terms of holographic principle \cite{Hol}. However, our current understandings of holographic principle are not complete as well.

\begin{figure}
  \centering
  \includegraphics[width=5cm]{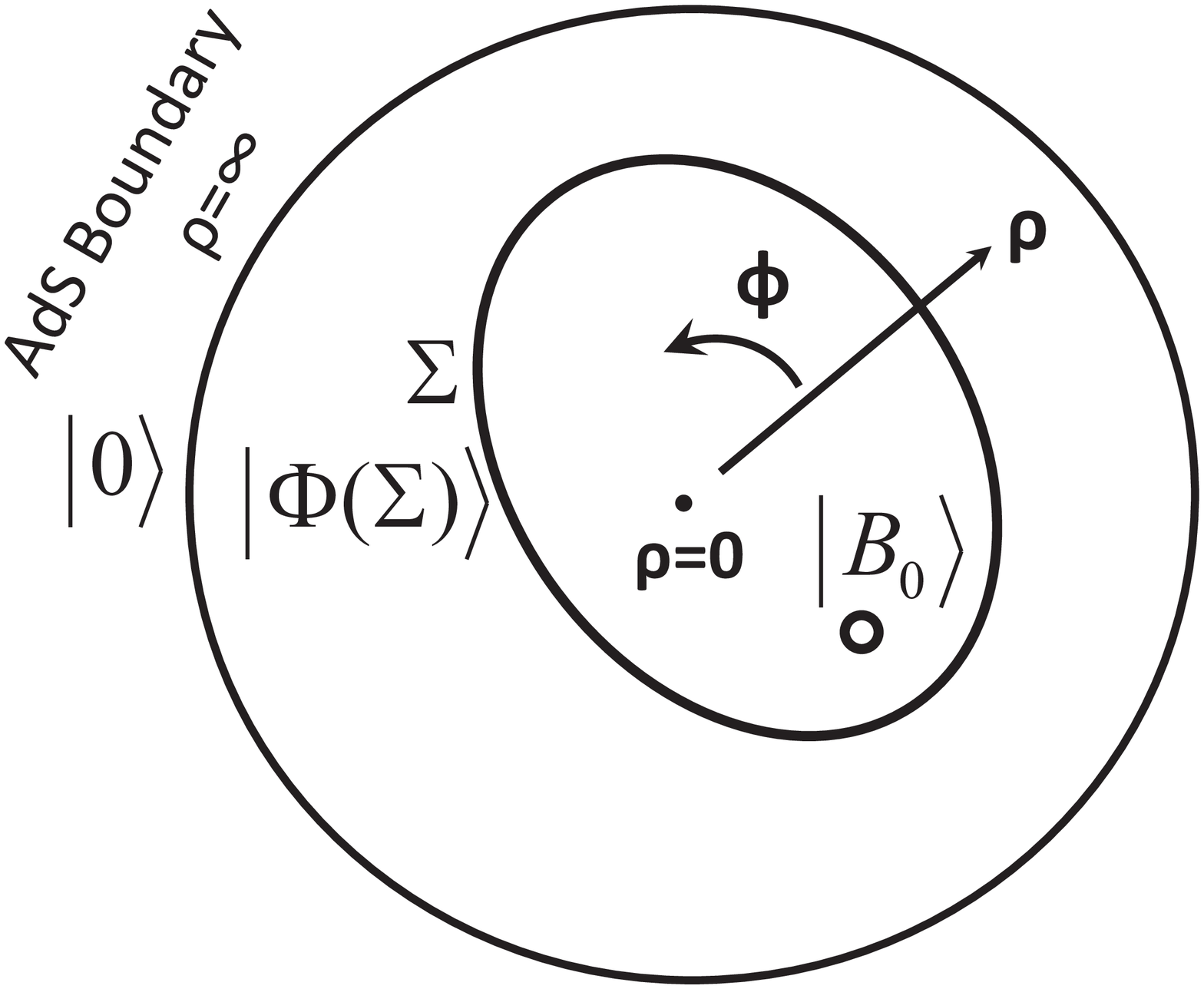}
  \caption{A sketch of Surface/State Correspondence in AdS/CFT.}
\label{fig:ss}
  \end{figure}

The recently proposed duality called surface/state correspondence \cite{MiTa} gives a more detailed structure of holographic relations. This duality can in principle be applied to any spacetimes described by Einstein gravity and even to those without time-like boundaries.
This surface/state correspondence (or simply called SS-duality) argues a correspondence between any codimension two convex surface $\Sigma$ and a quantum state described by a density matrix $\rho(\Sigma)$ for the Hilbert space of quantum theory dual to the Einstein gravity. When this surface is closed and topologically trivial, the state is given by a pure state $\rho(\Sigma)=|\Sigma\lb\la\Sigma|$. In particular, if we consider Einstein gravity in an AdS space and take $\Sigma$ to be a time slice of AdS boundary, then $|\Phi(\Sigma)\lb$ is simply given by the ground state $|0\lb$ of the dual conformal field theory (CFT). Refer to Fig.\ref{fig:ss}.

This SS-duality is argued based on the recently found connection between the AdS/CFT and the tensor networks. Such a relation has been first proposed in \cite{Swingle} for MERA (Multi-scale Entanglement Renormalization Ansatz) \cite{MERA} and later developed in \cite{NRT} for cMERA (continuous MERA) \cite{cMERA}. Refer also to e.g.\cite{Qi,PYHP,Bao,Cz} for various refinements and limitations of the connection between AdS/CFT and tensor networks. In general, a tensor network describes a wave function of a quantum state as a network diagram which fills a discretized space. The state $|\Phi(\Sigma)\lb$ dual to a convex closed surface $\Sigma$ is constructed by contracting the indices of tensors which are included in the region surrounded by $\Sigma$. For example, in the network found in \cite{PYHP} we can explicitly construct the state $|\Phi(\Sigma)\lb$ consistently by the above procedure.
If the tensor network describes correctly a CFT ground state, then we expect the space described by the network is
identical to a hyperbolic space, being equivalent to a time slice of AdS space. We would like to argue that the most direct way to realize tensor networks for CFTs is to employ the cMERA as we do not need to worry about lattice artifacts.

It is also important that the Hilbert space structure does not change under smooth deformation of $\Sigma$ in the SS-duality. Even though the discretized tensor network picture tells us that the Hilbert space for $|\Phi(\Sigma)\lb$ is given by the links of network intersecting with $\Sigma$ and thus its size can change, we always insert a dummy trivial state to keep the total dimension of Hilbert space the same. Thus the evolution of $|\Phi(\Sigma)\lb$ under a smooth deformation of $\Sigma$ can be treated as a unitary transformation.

The most elementary object in SS-duality is the quantum state dual to a zero size closed surface,
i.e. just a point. Such a state dual to a point in a gravitational spacetime is identified with the boundary state $|B\lb$ \cite{MiTa}. This is because there is no real space entanglement for the state dual to such a point-like surface, according to the idea of holographic entanglement entropy \cite{RT}, and because the state with a vanishing real space entanglement entropy is given by the boundary state \cite{MRTW}.

\begin{figure}
  \centering
  \includegraphics[width=5cm]{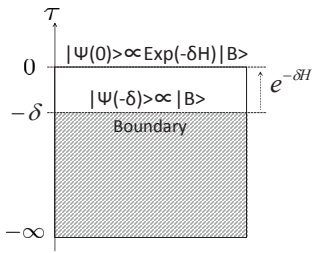}
  \caption{A realization of a boundary state in CFT by imposing a boundary condition in the path-integral formalism. As a UV regularization, we introduced an (Euclidean) time translation by the CFT Hamiltonian $H=L_0+\ti{L}_0-c/12$ for a short time interval $-\delta\leq \tau\leq 0$.}
\label{fig:bs}
  \end{figure}

The latter fact can be naturally
understood by turning off a relevant (e.g. mass) operator in a CFT suddenly at (Euclidean) time $\tau=0$ as in the analysis of quantum quenches \cite{cag}. In terms of quantum states we find that the ground state $|0\lb$ appears for $\tau>0$ and thus it is equivalent to put a sharp boundary at $\tau=0$ and restricts the spacetime for the region $\tau>0$ as in Fig.\ref{fig:bs}. In 2d CFTs, such a physical boundary state is called a Cardy states $|C_\ap\lb$ \cite{CS}, where $\ap$ labels the primary fields $\Psi_\ap$. An Ishibashi state $|I_\ap\lb$ \cite{Is} is a boundary state which includes only one sector of primary field $\Psi_\ap$ and its descendants.
A Cardy state is given by a specific linear combination of Ishibashi states.

The purpose of this paper is to give an explicit formulation of SS-duality for AdS/CFT correspondence by generalizing the formalism of cMERA. We would like to show how the bulk geometry appears and how the bulk operators are described in this formalism.
In particular we will focus on the setup of AdS$_3/$CFT$_2$ so that we have a good control of boundary states.\\

{\bf 2. SS-duality formulation and cMERA}

Let us start with a cMERA description of the CFT ground state $|0\lb$. We will employ the rescaled formalism in \cite{NRT} which is obtained from the original construction \cite{cMERA} by getting rid of the rescaling procedure and which has an advantage that the Hilbert space does not change even if we consider a CFT on a compact manifold. The cMERA formulation is defined by a flow from the UV state given by the CFT vacuum $|0\lb$ to the IR state which has no real space entanglement.

As we have explained, we can identify such a state with one of boundary states \cite{MRTW}, denoted by $|B_0\lb$.
In CFT$_{d+1}$ for $d\geq 3$, there is a conformal mass term in the gauge theory on $R\times S^d$ dual to a global AdS$_{d+2}$. Therefore it is natural to identify $|B_0\lb$ with the (Cardy-like physical) boundary state for Dirichlet boundary condition, which also preserves the R-symmety.  However, in a 2d CFT on a cylinder, there is no conformal mass and thus it is subtle whether $|B_0\lb$ is a Cardy state or Ishibashi state. Nevertheless, the requirement of preservation of R-symmetry and other internal symmetries suggests that the IR state should be the Ishibashi state $|I_0\lb$ for the identity primary state.

In this way, our general formulation of cMERA is expressed as follows:
\be
|0\lb={\cal P}\exp\left(-i\int^0_{-\infty}du \hat{K}(u)\right)|I_0\lb, \label{cMERAnew}
\ee
where $\hat{K}(u)$ is the disentangling operator at scale $u$ and ${\cal P}$ denotes the path-ordering. This operation $\hat{K}(u)$ eliminates quantum entanglement longer than the length scale $\ep e^{-u}$, where $\ep$ is the UV cut off or lattice spacing. The UV and IR limit corresponds to $u=0$ and $u=-\infty$, respectively.

Once $\hat{K}(u)$ is given, we can define the intermediate state $|0(u)\lb$  at scale $u$ as follows
\be
|0(u)\lb={\cal P}\exp\left(-i\int^u_{-\infty}du \hat{K}(u)\right)|I_0\lb. \label{cMERAnewp}
\ee
In this way, the cMERA is a unitary transformation (or a generalization of Bogoliubov transformation)
from the vacuum to the Ishibashi state. As
$u$ increases, some amount of quantum entanglement is added by the $\hat{K}(u)$ operation.
In the light of the AdS/CFT, we expect that the cMERA network describes the time slice of AdS space i.e. hyperbolic space.

In SS-duality, we can consider the surface $\Sigma(u)$ dual to
$|0(u)\lb$. This surface coincides with a time slice of the AdS boundary for $u=0$ and get shrinking as $u$ decreases. Eventually at $u\to -\infty$, it degenerates to a point at the origin of the AdS space. Therefore we have
$|0(0)\lb=|0\lb$ and $|0(-\infty)\lb=|I_0\lb$ as in Fig.\ref{fig:ss}.

Now we consider a 2d holographic CFT on a cylinder, whose coordinate is given by $(t,\phi)$ with the periodicity $\phi\sim \phi+2\pi$. The dual AdS$_3$ is given by the global coordinate
\be
ds^2=R^2(-\cosh^2\rho dt^2+d\rho^2+\sinh^2\rho d\phi^2). \label{ads3}
\ee
The isometry of AdS$_3$ is given by $SL(2,R)_L\times SL(2,R)_R$, which are generated by
 $(L_1,L_0,L_{-1})$ and $(\ti{L}_1,\ti{L}_0,\ti{L}_{-1})$ dual to the (global) Virasoro symmetry of 2d CFT. These are explicitly given by the following action in AdS$_3$ \cite{MaSt}:
\ba
&& L_0=i\de_+, \ \ \ti{L}_{0}=i\de_{-},  \no
&& L_{\pm 1}=ie^{\pm ix^+}\left[\f{\cosh2\rho}{\sinh2\rho}\de_{+}-\f{1}{\sinh2\rho}\de_{-}
\mp\f{i}{2}\de_\rho\right],\no
&& \ti{L}_{\pm 1}=ie^{\pm ix^-}\left[\f{\cosh2\rho}{\sinh2\rho}\de_{-}-\f{1}{\sinh2\rho}\de_{+}
\mp\f{i}{2}\de_\rho\right]. \label{Lth}
\ea

In particular, we are interested in a $SL(2,R)$ subgroup of $SL(2,R)_L\times SL(2,R)_R$ which does not change the time slice $t=0$. It is generated by $(l_1,l_0,l_{-1})$ defined by
\ba
&& l_0=L_0-\ti{L}_0=i\de_\phi, \no
&& l_{-1}=\ti{L}_{1}-L_{-1}=ie^{-i\phi}\left[-\f{1+\cosh(2\rho)}{\sinh(2\rho)}\de_\phi-i\de_\rho\right], \no
&& l_{1}=\ti{L}_{-1}-L_{1}=-ie^{i\phi}
\left[\f{1+\cosh(2\rho)}{\sinh(2\rho)}\de_\phi-i\de_\rho\right].
\ea
They satisfy the $SL(2,R)$ algebra as usual
\be
[l_0,l_{\pm 1}]=\mp l_{\pm 1},\ \ \ [l_{1},l_{-1}]=2l_0,
\ee
and correspond to the Killing vectors on $H_2$ defined by the time slice $t=0$ of the AdS$_3$.

The $SL(2,R)$ transformation $g(\rho,\phi)$ which takes the origin $\rho=0$ to
a point $(\rho,\phi)$ on $H_2$ is given by
\be
g(\rho,\phi)=e^{i\phi l_0}e^{\f{\rho}{2}(l_{1}-l_{-1})}.
\ee

It is obvious that the CFT vacuum $|0\lb$ is invariant under this $SL(2,R)$ transformation. Moreover, boundary states have the same invariance:
\be
g(\rho,\phi)|I_\ap\lb=|I_\ap\lb,  \label{syme}
\ee
 which comes from the basic property $(L_n-\ti{L}_{-n})|I_\ap\lb=0$ of the boundary states.
Thus the quantum states dual to points on the $H_2$ are all given by the same state $|I_0\lb$.
This agrees with the argument of SS-duality where all states whose dual surfaces are related by isometry are the same \cite{MiTa}. This is also consistent with the tensor network picture because this quantum state corresponds to the point-like state in the network and should be the same trivial state with no entanglement.

\begin{figure}
  \centering
  \includegraphics[width=5cm]{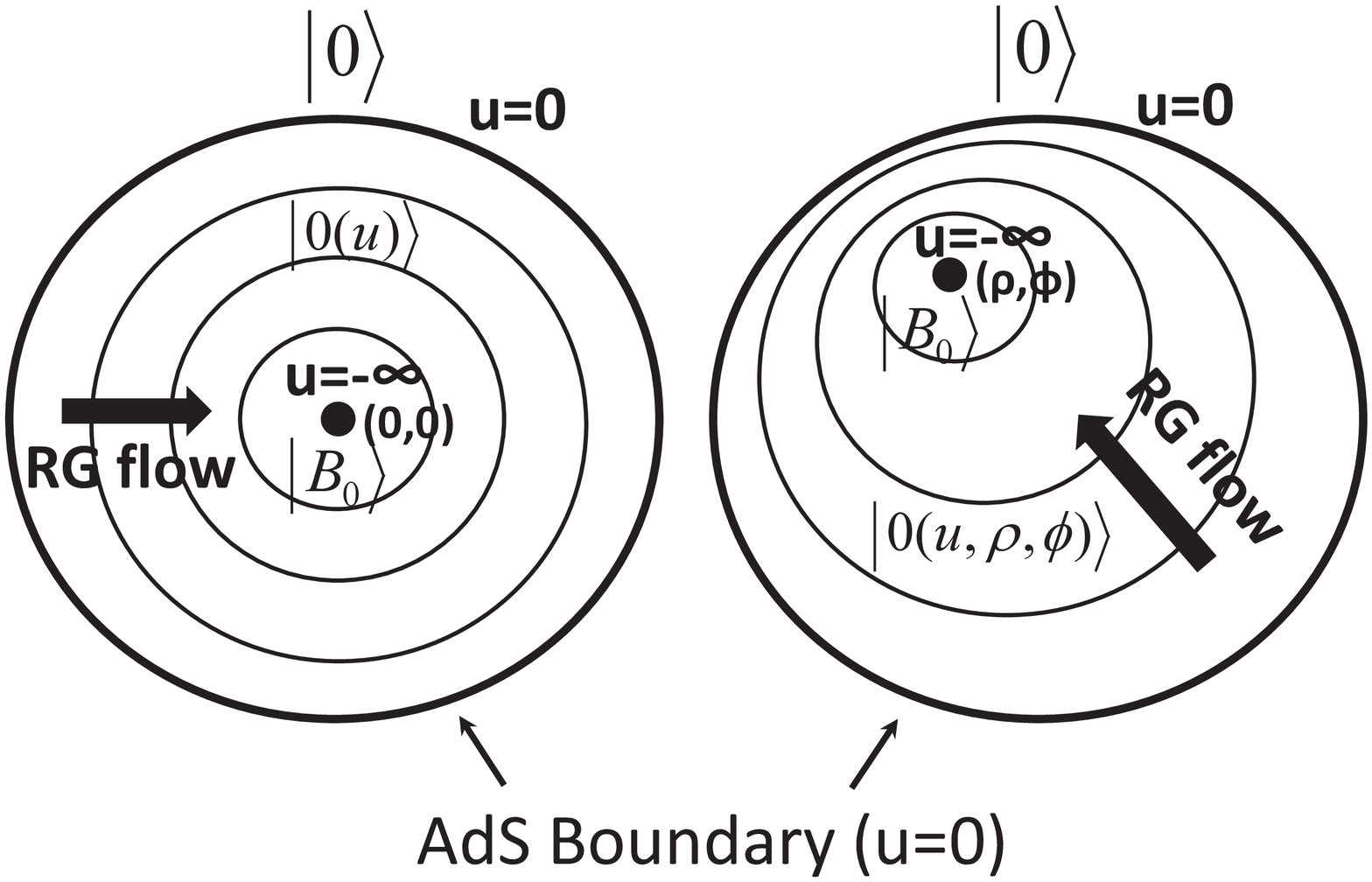}
  \caption{The $SL(2,R)$ conformal transformation of cMERA flow.}
\label{fig:cMERA}
  \end{figure}

By acting $g(\rho,\phi)$ transformation, we can rewrite (\ref{cMERAnew}) as
\be
|0\lb={\cal P}\exp\left(-i\int^0_{-\infty}du \hat{K}_{(\rho,\phi)}(u)\right)|I_0\lb, \label{cMERAneww}
\ee
where we defined
\be
\hat{K}_{(\rho,\phi)}(u)=g(\rho,\phi)\cdot \hat{K}(u)\cdot g(\rho,\phi)^{-1}. \label{sltwa}
\ee
This transformation relates two different the cMERA networks related by the conformal transformation as sketched in Fig.\ref{fig:cMERA}. For simplicity of our expressions, we introduce the notation:
\be
U_{(\rho,\phi)}= {\cal P}\exp\left(-i\int^0_{-\infty}du \hat{K}_{(\rho,\phi)}(u)\right).
\ee

We would also like to mention one more important observation. Since boundary states preserve $l_n=\ti{L}_{-n}-L_n$ even for $|n|\geq 2$, we can generalize (\ref{sltwa}) into
\be
\hat{K}_{h}(u)=\hat{h}(u)\hat{K}(u)\hat{h}(u)^{-1}+i\de_u \hat{h}(u)\cdot \hat{h}(u), \label{trsf}
\ee
where $\hat{h}(u)=\exp\left(\sum_{n}h_n(u)l_n\right)$. This transformation (\ref{trsf}) is interpreted as the deformation of the intermediate surface $\Sigma_u$ dual to the state
\be
|\Phi(\Sigma_u)\lb={\cal P}\exp\left(-i\int^u_{-\infty}du \hat{K}_g(u)\right)|I_0\lb, \label{cMERAnewkp}
\ee
which allows us to choose any possible foliation $\{\Sigma_u\}_{-\infty<u<0}$ of the time slice H$_2$.
Note that as long as we assume $h_n(0)=0$, we always end with up the vacuum state $|\Phi(\Sigma_0)\lb=|0\lb$ at $u=0$.
This confirms the proposed surface/state corresponence. At the same time, we find that the diffeomorphism gauge symmetry which preserves the time slice is included in our generalized cMERA formulation.\\

{\bf 3. Bulk Local Operators}

Now we would like to turn to excitations by bulk fields. There are two equivalent ways to realize this. One is to modify only the disentangler $\hat{K}(u)$ in (\ref{cMERAnew}).
For example, a construction of $\hat{K}(u)$ for quantum quenches have been given in \cite{NRT}. Another is to modify only the IR state $|I_0\lb$, keeping the cMERA network unchanged. Analogous formulation was constructed for lattice models in \cite{Qi,PYHP}. Below we will work in the latter formalism and thus $\hat{K}(u)$ is the same as the one for the ground state.

In AdS/CFT, the vacuum $|0\lb \in H_{CFT}$ of the CFT is dual to the vacuum of
bulk gravity $|0\lb_{bulk}\in H_{bulk}$, where $H_{CFT}$ and $H_{bulk}$ denote the Hilbert space of the CFT and the bulk gravity, respectively.

 Let us insert a bulk quantized field $\hat{\psi}_\ap(\rho,\phi)$ dual to a CFT primary field $\Psi_\ap$, on the time slice $H_2$ at the point $(\rho,\phi)$. We argue that the locally excited state $\hat{\psi}_\ap(\rho,\phi)|0\lb_{bulk}$ is dual to the following CFT excited state $|\Psi_\ap(\rho,\phi)\lb$:
\ba
&&\hat{\psi}_\ap(\rho,\phi)|0\lb_{bulk}\in H_{bulk} \no
&&\ \ \ \lr\  |\Psi_\ap(\rho,\phi)\lb=U_{(\rho,\phi)}|I_\ap\lb \in H_{CFT}. \label{dualex}
\ea
Since the metric should not change for the locally excited state except the localized region, the state should have almost vanishing real space entanglement. Therefore, the IR state of the cMERA for the excited is also given by a boundary state. Since the primary field $\Psi_\ap$ is included only in $|I_\ap\lb$, we can argue that the IR state is given by $|I_\ap\lb$ as in (\ref{dualex}).

By taking into the time evolution, we can make the following identification
\ba
\hat{\psi}_\ap(\rho,\phi,t) \lr e^{i(L_0+\ti{L}_0)t}\cdot U_{(\rho,\phi)}\cdot M_\ap\cdot  U_{(\rho,\phi)}^{-1}\cdot e^{-i(L_0+\ti{L}_0)t}, \label{dualexx}\no
\ea
where $M_\ap$ is an unitary operation which maps $|I_0\lb$ to $|I_\ap\lb$. Refer to e.g. \cite{Kabat} for an earlier standard literature of
bulk field construction from CFT operators. We are taking a different route for the same purpose as we are restricting on a specific time slice instead of working in a Lorentz covariant formalism.

It is obvious that we can take any linear combinations of them as
\ba
&& \int_{H_2} d\rho d\phi\  f_\ap(\rho,\phi) \hat{\psi}_\ap(\rho,\phi)|0\lb_{bulk}\in H_{bulk}\no
&&\ \ \ \lr\  \int_{H_2} d\rho d\phi\  f_\ap(\rho,\phi) |\Psi_\ap(\rho,\phi)\lb \in H_{CFT}.
\ea
Moreover we can consider multiply excited states
\ba
&& \hat{\psi}_\ap(\rho,\phi)\hat{\psi}_\beta(\rho',\phi')|0\lb_{bulk}\in H_{bulk}\no
&& \lr\ (U_{(\rho,\phi)}\cdot M_\ap\cdot  U_{(\rho,\phi)}^{-1})(U_{(\rho',\phi')}\cdot M_\beta\cdot  U_{(\rho',\phi')}^{-1})|0\lb \in H_{CFT}.\nonumber
\ea

In this way, we find the correspondence
\be
|\psi_{bulk}\lb\in H_{bulk}=\otimes_{(\rho,\phi)}H_{bulk}^{(\rho,\phi)}\ \to\  |\psi\lb\in H_{CFT}.
\ee
However this map should not be an equivalent relation as the size of boundary CFT Hilbert space is limited. Indeed, we expect that the bulk field theory should have the UV cut off at the Planck scale $l_p$.

We can estimate how much the locality of bulk local operators persists when the length of $(\rho'-\rho,\phi'-\phi)$ gets larger by looking at the inner product
\be
\la I_\ap|U_{(\rho,\phi)}^{-1}U_{(\rho',\phi')}|I_\ap\lb=\la \Psi_\ap(\rho,\phi)|\Psi_\ap(\rho',\phi')\lb, \label{inpw}
\ee
where we used the symmetry (\ref{syme}). If this inner product is much smaller than one, then we expect that the bulk fields behave locally for that length scale.\\

{\bf 4. Scalar Field Wave Functions}

Next we need to understand what the states $|\Psi_\ap(\rho,\phi)\lb$ look like. For our convenience, let us define
$|\Psi_\ap\lb\equiv |\Psi_\ap(0,0)\lb$ such that $|\Psi_\ap(\rho,\phi)\lb=g(\rho,\phi)|\Psi_\ap\lb$.  Remember that $|\Psi_\ap\lb$ is a CFT excited state which is dual to a local excitation in the bulk AdS at the origin $\rho=0$ at the time $t=0$. Thus we are considering a bulk geometry where only the origin of AdS$_3$ at $t=0$ differs from that of pure AdS$_3$.

 The $SL(2,R)$ subgroup of the original
 $SL(2,R)_L\times SL(2,R)_R$ which preserves the point $\rho=t=0$ is not same as the one generated by $l_0,l_{\pm}$, but it is generated by $L_0-\ti{L}_0,\ L_{1}+\ti{L}_{-1}$ and $L_{-1}+\ti{L}_{1}$, as can be confirmed from (\ref{Lth}). Therefore $|\Psi_\ap\lb$ should satisfy
 \be
 (L_0-\ti{L}_0)|\Psi_\ap\lb=(L_{1}+\ti{L}_{-1})|\Psi_\ap\lb=(L_{-1}+\ti{L}_{1})|\Psi_\ap\lb=0.   \label{hwsl}
 \ee
  The simplest solution to this condition (\ref{hwsl}) is given by
 \be
|\Psi_\ap\lb \propto  e^{i\f{\pi}{2}(L_0+\ti{L}_0)}|J_\ap\lb,  \label{exjkjfrg}
 \ee
where $|J_\ap\lb$ is the `Ishibashi state' for the $SL(2,R)_L\times SL(2,R)_R$ subalgebra of the Virasoro algebra.
It is explicitly defined by
\be
|J_\ap\lb=\sum_{k=0}^\infty |k\lb_L |k\lb_R,
\ee
where $|k\lb_L$ (or $|k\lb_R$) denotes the normalized (unit norm) descendant state proportional to
$(L_{-1})^k|\ap\lb$ (or $(\ti{L}_{-1})^k|\ap\lb$).
We would like to argue that this choice (\ref{exjkjfrg}) is the correct state dual to the bulk local operator.
Note that if we in particular choose the primary state $|\ap\lb$ to be the vacuum $|0\lb$, we simply find
$|\Psi_\ap(\rho,\phi)\lb=|0\lb$ as expected.

Indeed, by using the property (\ref{hwsl}), we can reproduce the correct scalar field solution on the AdS$_3$ dual
to the state $|\beta\lb$, which is either a primary state $|\ap\lb$ or one of its descendants. The bulk scalar field expectation value for the state $|\beta\lb$ can be computed from the CFT inner product as follows
\ba
\la\hat{\psi}_\ap(\rho,\phi,t)\lb_{|\beta\lb}=\la \Psi_\ap(\rho,\phi)|e^{-it(L_0+\ti{L}_0)}|\beta\lb.
\label{wert}
\ea

By using the identity
\ba
\la \Psi_\ap|e^{-\f{\rho}{2}(l_{1}-l_{-1})}|\ap\lb=\la \Psi_\ap|e^{\rho(L_{1}-L_{-1})}|\ap\lb =\f{1}{(\cosh\rho)^{2\Delta_\ap}},\nonumber
\ea
we can confirm that the scalar field expectation value for the primary state agrees with the known scalar field
solution in AdS$_3$:
\be
\la\hat{\psi}_\ap(\rho,\phi,t)\lb_{|\ap\lb}\propto e^{-2i\Delta_\ap t}\f{1}{(\cosh\rho)^{2\Delta_\ap}},\label{wfed}
\ee
where $(L_0-\Delta_\ap)|\ap\lb=(\ti{L}_0-\Delta_\ap)|\ap\lb=0$.

We can also extend the matching of (\ref{wert}) with those of scalar field solutions on the global AdS$_3$
to SL(2,R) descendants states as we will show explicitly in the appendix A.
They are obtained by acting $L_{\pm 1}$ and
$\ti{L}_{\pm 1}$ on the primary state and the scalar field expectation values are obtained by acting
 the differential operators (\ref{Lth}) on (\ref{wfed}). It is also useful to note the inner product (\ref{wert}) is vanishing if we choose $|\beta\lb$ to be non-$SL(2,R)$ descendants which are orthogonal to $|k\lb_L$ and $|k\lb_R$.
 Thus we can confirm that the (perturbative) equation of motion for the scalar field operator
\be
\left[L^2+\ti{L}^2+\f{m^2R^2}{2}\right]\hat{\psi}_\ap(t,\rho,\phi)=0,
\ee
is satisfied. Here $L^2=(L_{-1}L_1+L_1L_{-1})/2-L_0^2$ is the differential operator corresponding to the Casimir of $SL(2,R)$ in terms of (\ref{Lth}) and we have  $\Delta_\ap=\f{1}{2}+\f{1}{2}\s{m^2R^2+1}$ as usual in AdS$_3/$CFT$_2$. Moreover, as we will show in the appendix B, we can prove that our inner product $\la \Phi(\rho,t,\phi) | \Phi(\rho',t,\phi') \lb$ in the 2d CFT perfectly matches with the known expression of bulk to bulk propagator of a free massive scalar in AdS$_3$.

In this way, we learned that the effect of unitary transformation $U_{(\rho,\phi)}$ is to remove the (higher) Virasoro generators $L_n$ and $\ti{L}_{n}$ with $|n|\geq 2$ and perform a time translation by $\f{\pi}{2}$.\\

{\bf 5. Information Metric}

To study the behavior of the inner product (\ref{inpw}), it is useful to employ the idea of Fisher information metric $G_{ab}$ defined by
\ba
&& 1-|\la \Psi_\ap(\rho,\phi)|\Psi_\ap(\rho+d\rho,\phi+d\phi)\lb| \no
&& =G_{\rho\rho}d\rho^2+2G_{\rho\phi}d\rho d\phi+G_{\phi\phi}d\phi^2.
\ea
This measures the distance between the two states $|\Psi_\ap(\rho,\phi)\lb$ and
$|\Psi_\ap(\rho+d\rho,\phi+d\phi)\lb$. In this way, we want to probe the AdS geometry by considering the distance
between two local excitations.

However, for our purpose, we do not want to have a literally delta functional local excitation but want to smear over a length larger than the Planck length near the origin. Indeed the state (\ref{exjkjfrg}) is singular in that it has an infinite norm and thus we need a UV regularization. We expect that the energy of local excitation should not exceed the Planck energy. In AdS/CFT, the energy $E$ is related to the conformal dimension $\Delta$ of operator via $E=\Delta/R$. If we substitute $E\ll 1/l_p\sim 1/G_N$, then we have the bound $\Delta\ll c$. Therefore we would like to argue the following
estimation:
\be
|\Psi_\ap\lb\propto e^{-\delta(L_0+\ti{L}_0)}e^{i\f{\pi}{2}(L_0+\ti{L}_0)}|J_\ap\lb,
\ee
where $\delta$ provided the UV cut off and therefore we should take $\delta\sim \f{1}{c}$.

We can evaluate the inner product keeping only quadratic terms:
\ba
&& 1-|\la \Psi_\ap(\rho,\phi)|\Psi_\ap(\rho+d\rho,\phi+d\phi)\lb| \no
&&=\f{1}{8}\left(d\rho^2+\sinh^2\rho d\phi^2\right)
\la \Psi_\ap|(l_{-1}l_1+l_1l_{-1})|\Psi_\ap\lb.\label{metcom}
\ea
Here we employed (\ref{syme}) and the identity
\be
e^{-\f{\rho}{2}(l_{1}-l_{-1})}l_0e^{\f{\rho}{2}(l_{1}-l_{-1})}
=\cosh\rho\cdot l_0-\sinh\rho\cdot \f{l_1+l_{-1}}{2}.
\ee
Also we can easily show
\ba
 l_{\pm 1}|\Psi_\ap\lb=-(e^{\pm 2\delta}+1)L_{\pm 1}|\Psi_\ap\lb. \label{lind}
\ea
Since the following identity is obvious
\be
\la \Psi_\ap|l_{-1}l_1|\Psi_\ap\lb=\la \Psi_\ap|l_{1}l_{-1}|\Psi_\ap\lb,
\ee
we can combining this with (\ref{lind}) to find
\be
2\delta\cdot \la \Psi_\ap|L_{-1}L_1|\Psi_\ap\lb=\la \Psi_\ap|L_0|\Psi_\ap\lb. \label{lopw}
\ee

This allows us to estimate the right hand side of (\ref{lopw}) by taking a derivative of $\delta$:
\ba
\la \Psi_\ap|L_0|\Psi_\ap\lb \simeq -\f{1}{4}\f{\de}{\de\delta}\left[\log \sum_{k=0}^\infty e^{-4\delta k}\right]\simeq  \f{1}{4\delta},
\ea
and therefore we obtain
\be
\la \Psi_\ap|l_{-1}l_1|\Psi_\ap\lb\simeq 4\la \Psi_\ap|L_{-1}L_1|\Psi_\ap\lb\simeq\f{1}{2\delta^2}.
\ee

In this way, the information metric for $|\Psi_\ap(\rho,\phi)\lb$ is given by that of a hyperbolic space:
\be
ds_{inf}^2=\f{1}{8\delta^2}\left(d\rho^2+\sinh^2\rho d\phi^2\right). \label{bsinf}
\ee
It is natural to expect that this corresponds to the time slice of the global AdS$_3$, to which our 2d CFT is dual. Indeed, the radius of this $H_2$ coincides with that in the AdS$_3$ metric with the Planck unit
up to an $O(1)$ numerical factor as we chose $\delta\sim 1/c$.

Actually, it is not difficult to obtain the full spacetime metric of AdS$_3$ including the time components. As we have shown in the appendix B, the
 two point function $\la \Phi(\rho,t,\phi) | \Phi(\rho',t,\phi') \lb$ coincides with the bulk to bulk propagator of a free massive scalar. Even more generally, when two points $X$ and $Y$ are closed to each other, any two point function of a $d+1$ dimensional free scalar field between them gets proportional to $D(X,Y)^{-(d-1)}$,
 where $D(X,Y)$ is the distance between the two points. If we regularize this by introducing a cut off $\delta$, then the corresponding normalized inner product looks like
 \be
 \la \Phi(X)|\Phi(Y)\lb\simeq \f{\delta^{d-1}}{\left(D(X,Y)^2+\delta^2\right)^{\f{d-1}{2}}}.
 \ee
By expanding w.r.t the infinitesimally small distance, we obtain
$ds^2_{inf}\propto \f{1}{\delta^2}g_{ij}dX^i dX^j$, where $g_{ij}$ is the metric of the bulk spacetime. Since it is natural to choose $\delta$ to be the Planck scale, the information metric $ds^2_{inf}$ coincides with the bulk metric in the Planck unit.\\

{\bf 6. Conclusions}

In this article, we gave an explicit construction of the conjectured surface/state correspondence in the setup of AdS$_3/$CFT$_2$. We realized this construction by generalizing the formulation of cMERA, where the boundary states in conformal field theories played an important role of describing points in AdS$_3$. Our formalism naturally
takes into account the bulk diffeomorphism as a gauge symmetry of cMERA formalism. We found an identification of bulk local operators which reproduces solutions of scalar field equations of motion on AdS$_3$. We also computed the information metric for a locally excited state and showed that it is given by that of a 2d hyperbolic manifold, which
is argued to describe the time slice of AdS$_3$.

An obvious and important future problem is to find the expression of disentangler $\hat{K}(u)$ for holographic CFTs. It is natural to expect that in the UV region $u\simeq 0$, $\hat{K}(u)$ gets qualitatively similar to that for free CFTs, where $\hat{K}(u)$ is a bilinear of creation and annihilation operators, which add $O(c)$ entanglement, as in \cite{cMERA,NRT}. On the other hand, we expect that in the IR region $|u|\ll 1$, the disentangler $\hat{K}(u)$ should be a linear combination of products of particular singlet operators such as $L_{-n}\ti{L}_{-n}$, which add only $O(1)$ entanglement in the IR, motivated by the confinement/deconfinement transition in holographic CFTs. It is curious to note that this IR behavior may suggest cMERA can have a sub-AdS scale locality, as opposed to (discrete) MERA.

After we finish this work, we noticed the interesting paper \cite{Ver}, which gave an identification of bulk local operator using boundary states. Even though its connection to our present
construction is not immediately clear, it might be possible to relate them by using the relation \cite{EV} which connects the path-integral on the CFT to the tensor network which describes the time slice of AdS$_3$.
\\

{\bf Acknowledgements} We thank John Cardy, Bartlomiej Czech, Sumit Das, Matthew Headrick, Sunil Mukhi, Masahiro Nozaki, Hirosi Ooguri, Xiao-Liang Qi, Shinsei Ryu, James Sully, Erik Tonni, Sandip Trivedi, Herman Verlinde, Guifre Vidal, and Beni Yoshida for useful conversations. We are grateful for stimulating discussions to the organizers and participants of ``International Workshop on Condensed Matter Physics and AdS/CFT'' held in Kavli IPMU, Tokyo University and the conference ``Closing the entanglement gap: Quantum information, quantum matter, and quantum fields'' held in KITP, UCSB, where this work has been completed. TN, NS and KW are supported by JSPS fellowships. TT is supported by JSPS Grant-in-Aid for Scientific Research (B) No.25287058. TT is also supported by World Premier International Research Center Initiative (WPI Initiative) from the
Japan Ministry of Education, Culture, Sports, Science and Technology
(MEXT).\\

\appendix

{\bf Appendix A: Analysis of Scalar Wave Functions}

We are interested in the inner product ($x^\pm\equiv t\pm\phi$)
\ba
G_{|\beta\lb}(\rho,t,\phi) &\equiv & \la \Psi_{\ap} | e^{\rho (L_1 - L_{-1})} e^{ - i \phi (L_0 - \bar{L}_0)} e^{-i(L_0+\bar{L}_0) t} |\beta \lb \no
&=&  \la \Psi_\ap(t) | e^{\rho(e^{-ix^+}L_1-e^{ix^+}L_{-1})}|\beta\lb. \label{fung}
\ea

Here $|\Psi_\ap(t)\lb$ takes the form
\be
|\Psi_\ap(t)\lb=\sum_{k=0}^\infty e^{i(\Delta_\ap+k)\pi} e^{2i(\Delta_\ap+k)t}|k\lb_L |k\lb_R. \label{ewq}
\ee

We assume the state $|\beta\lb$ has the factorized form
\be
|\beta\lb=|k\lb_L|\bar{k}\lb_R,
\ee
where the conformal dimensions are given by $L_0=\Delta_{\ap}+k$ and $\ti{L}_0=\Delta_{\ap}+\bar{k}$, which can be different from each other. It is obvious that $G_{0}(\rho,t,\phi)\propto e^{-i(\Delta_{\ap}+k)x^+-i(\Delta_{\ap}+\bar{k})x^-}$.

Now we would like to find a relation between
\be
G_{L_{-1}|\beta\lb}(\rho,t,\phi)\equiv \la \Psi_\ap(t) | e^{\rho(e^{-ix^+}L_1-e^{ix^+}L_{-1})}L_{-1}|\beta\lb,
\ee
and $G_{|\beta\lb}(\rho,t,\phi)$ in (\ref{fung}).

To see this, we consider the following inner product:
\be
\la \Psi_\ap(t) | L_0 e^{\rho(e^{-ix^+}L_1-e^{ix^+}L_{-1})}|\beta \lb
=(\Delta_\ap+\bar{k}) G_{|\beta\lb}(\rho,t,\phi).  \label{ewqz}
\ee

We can rewrite this as follows:
\ba
&& (\Delta_\ap+\bar{k}) G_{|\beta\lb}(\rho,t,\phi) \no
&& =\la \Psi_\ap(t) | L_0 e^{\rho(e^{-ix^+}L_1-e^{ix^+}L_{-1})} |\beta \lb  \no
&&=(\Delta_\ap+k)\cosh2\rho\cdot G_{|\beta\lb}(\rho,t,\phi)-\f{\sinh2\rho}{2}\de_\rho G_{|\beta\lb}(\rho,t,\phi)\no
&&\ \ \ \ -e^{ix_+}\sinh 2\rho\cdot G_{L_{-1}|\beta\lb}(\rho,t,\phi),
\ea
where we employed the identity
\ba
&& e^{-\rho(e^{-ix^+}L_1-e^{ix^+}L_{-1})}L_0e^{\rho(e^{-ix^+}L_1-e^{ix^+}L_{-1})} \no
&& =\cosh2\rho \cdot L_0-\f{\sinh 2\rho}{2}(e^{-ix^+}L_1+e^{ix^+}L_{-1}).
\ea
We also used the decomposition
\be
e^{-ix^+}L_1+e^{ix^+}L_{-1}=(e^{-ix^+}L_1-e^{ix^+}L_{-1})+2e^{ix^+}L_{-1}, \nonumber
\ee
where the first term in the right-hand side is equivalent to the derivative $\de_\rho$.

Thus by equating the first and last equation in the above we obtain
\ba
&& G_{L_{-1}|\beta\lb}(\rho,t,\phi) \no
&& =ie^{-ix_+}\left(\f{i}{2}\de_\rho-\f{1}{\sinh2\rho}\de_{x^-}
+\f{\cosh2\rho}{\sinh2\rho}\de_{x^+}\right)G_{|\beta\lb}(\rho,t,\phi).
\no \label{equiv}
\ea
This indeed coincides with the differential operator $L_{-1}$.  We can obtain a similar proof for
$L_1$, $\ti{L}_{\pm}$.\\

{\bf Appendix B Analysis of Two Point Functions}

In this section we show that the inner product $\la \Phi(\rho,t,\phi) | \Phi(\rho',t,\phi') \lb$ reproduces the bulk to bulk propagator of a free massive scalar in AdS$_3$.

The solutions of equations of motion $(\Box_{AdS3}+m^2)\Phi=0$ for a free massive scalar $\Phi$ which correspond to general descendant states in CFT$_2$ are given
in terms of Jacobi Polynomials:
(see eq.(28) in \cite{Bulk})
\ba
&& %%\la\Phi(\rho,t,\phi)|k\lb _L |\bar{k}\lb_R =
\Phi_{k,\bar{k}}(\rho,t,\phi) \no
&=&e^{-i(2\Delta+k+\bar{k})t}e^{i(k-\bar{k})\phi}(\sin\theta)^{|k-\bar{k}|}(\cos\theta)^{2\Delta} \no
&&\ \ \ \times P^{(|k-\bar{k}|,2 \Delta -1)}_{\bar{k}}(\cos 2\theta).
\ea
where we introduce the coordinate $\theta$ via $\sin \theta = \tanh^{-1} \rho$; the Jacobi polynomials $P^{(\ap,\beta)}_n(x)$  (which is a generalization of Legendre function) are defined using the hypergeometric function:
\be
P^{(\ap,\beta)}_n(x)\equiv \f{(n+\ap)!}{n!\ap!}\cdot~ {}_2F_1\left(\!-n,n+\ap+\beta+1,\ap+1;\f{1-x}{2}\right)
\ee

By using the $SL(2,R)$ invariance, we can set $\rho'=\phi'=t'=0$ in the two point function without losing generality:
\ba
 G(\rho,t)
 &=&\la \Psi(\rho,\phi,t)|\Psi(0,0,0)\lb \no
 &=& \la \Psi_\ap(t)|e^{\rho (e^{-ix^+}L_1-e^{ix^+}L_{-1})}|\Psi_\ap\lb,\no
 &=& \la \Psi_\ap(t)|e^{\rho (L_1-L_{-1})}|\Psi_\ap\lb.
  \ea
 This clearly shows that it is $\phi$ independent as expected from the rotational invariance.

 We can insert the identity
\be
1=\sum^\infty_{k,\bar{k}=0}|k\lb_L |\bar{k}\lb_R \la k|_L \la \bar{k}|_R.
\ee
in the above two point function. The state $|k\lb_{L,R}$ is explicitly given by
\be
|k\lb_L=\f{1}{\s{N_k}}(L_{-1})^k|\ap\lb,\ \ \ \  |k\lb_R=\f{1}{\s{N_k}}(\ti{L}_{-1})^k|\ap\lb,
\ee
where
\be
N_k=\la \ap|(L_1)^k(L_{-1})^k|\ap\lb=\prod_{j=1}^k (j^2+(2\Delta-1)j).
\ee
Then, we find
\ba
G(\rho,t)&=&\la \Psi_\ap(t)|e^{\rho (L_1-L_{-1})}
\left(\sum^\infty_{k,\bar{k}=0}|k\lb_L |\bar{k}\lb_R \la k|_L \la \bar{k}|_R\right)|\Psi_\ap\lb \no
&=&\sum_{k=0}^\infty (-1)^k  \la \Psi_\ap(t)|e^{\rho (e^{-ix^+}L_1-e^{ix^+}L_{-1})}|k\lb_L |k\lb_R,
\ea
where we employed (\ref{ewqz}) at $t=0$.

By using the result (\ref{equiv}) on the equivalence between the $L_{-1},\ti{L}_{-1}$ action on the state
and its differential action on AdS$_3$ in \cite{MaSt} and
the following formula for Jacobi polynomials
\ba
&&(1-x^2) \f{d P^{(\ap,\beta)}_{n+1}(x)}{dx} +( n+1)(x+\f{\beta - \ap}{2n +\ap+\beta + 2} ) P^{(\ap,\beta)}_{n+1}(x)\no
 &&= \f{2(n+\ap+1)(n+\beta + 1)}{2n+\ap+\beta + 2}P^{(\ap,\beta)}_n(x) \no
 &&-(1-x^2) \f{d P^{(\ap,\beta)}_n(x)}{dx} \no
  && \ \ \ \ \ \ \ \ \ \  +( n+\ap + \beta +1)(x+\f{  \ap -\beta}{2n +\ap+\beta + 2} ) P^{(\ap,\beta)}_n(x)\no
  &&= \f{2(n+1)(n+\ap+\beta + 1)}{2n+\ap+\beta + 2}P^{(\ap,\beta)}_{n+1}(x),
\ea
we find
\be
 L_{-1}\ti{L}_{-1}\Phi_{k,k} \no
=-(k+1)(k+2\Delta)\Phi_{k+1,k+1}.
\ee
Using this relation recursively, we find
\ba
\la \Psi_\ap(t)|e^{\rho (e^{-ix^+}L_1-e^{ix^+}L_{-1})}|k\lb_L |k\lb_R&=&\f{1}{N_{k}}(L_{-1})^k(\ti{L}_{-1})^k \Phi_{0,0} \no
&=&(-1)^k \Phi_{k,k}.
\ea

In this way we can evaluate the two point function as follows:
\ba
G(\rho,t)&=&\sum_{k=0}^\infty  \Phi_{k,k}(\rho,t)\no
&=&\sum_{k=0}^\infty \f{e^{-i2(k+\Delta)t}}{(\cosh\rho)^{2\Delta}}P_{k}^{(0,2\Delta-1)}(1-2\tanh^2\rho).
\ea
By using the formula for generating functions of Jacobi polynomials
\be
\f{2^{\alpha+\beta}}{R(1+R-z)^{\alpha}(1+R+z)^{\beta}}=\sum_{n=0}^\infty P_{n}^{(\alpha,\beta)}(x)z^n,  \label{qq}
\ee
where we assume $R = \sqrt{1 - 2 xz + z^2 }$, we can show
\be
G(\rho,t) = \f{e^{-(2\Delta-1)D(\rho,t)}}{2 \sinh D(\rho,t)}.
\ee
Here the geodesic length $D(\rho,t)$ in AdS$_3$ is given by
\be
\cosh D(\rho,t) = \cosh \rho \cos t.
\ee
This precisely reproduce the known expression of the Green function (or bulk to bulk propagator) for arbitrary conformal dimension $\Delta$
(see e.g. \cite{BCFM,Green}).

\end{document}